\documentstyle[12pt,epsf]{article}
\newcommand{\lsim}{\,{\buildrel < \over {_\sim}}\,}
\newcommand{\gsim}{\,{\buildrel > \over {_\sim}}\,}

\begin{document}

\begin{titlepage}
\begin{flushright}
CERN-TH/97-220\\
hep-ph/9708472\\
\end{flushright}
\begin{centering}
\vfill

{\bf 
Pre-thermalization dynamics: initial conditions for QGP \\
at the LHC and RHIC from perturbative QCD
\footnote{a lecture given at the ``International 
School on the Physics of Quark Gluon Plasma'', June 3-6, 1997, Hiroshima;
to appear in Prog. Theor. Phys. (Proc. Suppl.)}
}
 
\vspace{0.5cm}
 K.J. Eskola\footnote{kari.eskola@cern.ch}

\vspace{1cm}
{\em CERN, Theory Division, \\CH-1211 Geneva 23, Switzerland\\}
and\\
{\em Department of Physics, University of Jyv\"askyl\"a,\\
P.O. Box 35,  FIN-40351 Jyv\"askyl\"a, Finland\\}
 
\vspace{1cm}
{\bf Abstract}

\end{centering}

\vspace{0.3cm}\noindent
I discuss how the initial conditions for QGP-production in 
ultrarelativistic heavy ion  collisions at the LHC and RHIC can 
be computed from perturbative QCD.
 
\vfill 
\noindent
CERN-TH/97-220\\
August 1997

\end{titlepage}

\section{Introduction}

Particle and transverse energy production in the central rapidity 
region of ultrarelativistic heavy ion collisions (URHIC) can be 
viewed as a combination of perturbative  (hard and semihard) parton 
production and  non-perturbative (soft) particle production.  
The key  feature in the division is that one should be able to 
{\it compute}  the hard and semihard parton production from 
perturbative QCD (pQCD), provided that the parton distributions 
of colliding nuclei are known \cite{BM87,KLL,EKL}. On top of this, 
the non-perturbative particle  production can be modelled  {\it e.g.}  
through strings \cite{STRINGS,HIJING}, perhaps also through a decaying 
strong  background  colour field  \cite{KM85}. 

The semihard QCD-processes are  expected to become increasingly important 
with increasing cms-energies $\sqrt s$, particularly in URHIC, due to the 
following reasons: 
Firstly, already in $p\bar p(p)$ collisions the  rapid rise of the total 
and inelastic cross sections  can be explained by copious production of 
semihard partons, ``minijets,'' with transverse momenta
$p_{\rm T}\ge p_0\sim 1...2$ GeV \cite{EIKONAL}. In other words, the events 
tend to become more ``minijetty'' towards higher $\sqrt s$. This is also  
expected to happen in $AA$ collisions, and even more so, 
because of the $A^{4/3}$-scaling of hard collisions.

Secondly, the rapid rise of the structure function $F_2$ observed at HERA
at $x\lsim0.01$ and $Q\gsim1$ GeV \cite{HERA} 
implies that minijet production at the LHC gets a further boost \cite{EKR}:
the  mid-rapidity minijets with  $p_{\rm T}\sim$ 2 GeV typically probe the 
gluon distributions at $x\sim 2p_{\rm T}/\sqrt s\sim 7\times 10^{-4}$ for 
nuclear collisions at $\sqrt s/A = 5.5$ TeV.

Thirdly, the time scale for producing partons  and transverse energy into 
the  central rapidity region by semihard collisions is short, typically  
$\tau_{\rm h}\sim 1/p_0\sim 0.1$ fm$/c$, where $p_0\sim 2$ GeV 
is the smallest  transverse momentum included in the computation. 
The soft processes are completed  at later stages of the collision, 
typically at $\tau_{\rm s}\sim 1/\Lambda_{\rm QCD}\sim 1$ fm$/c$. 
If the density  of partons produced in the hard and semihard stages 
of the heavy ion collision becomes high enough - as will be the case 
- a saturation in the initial parton production can occur already in the 
perturbative region $p_{\rm T}\gg\Lambda_{\rm QCD}$ \cite{BM87,GLRpr,EMW,EK96}, 
and, consequently, softer particle production will be screened. 
The fortunate consequence of this is that a larger part of 
parton  production in the central rapidities can be {\it computed} 
from perturbative QCD (pQCD) at higher energies, and the relative 
contribution from soft collisions with $p_{\rm T}\lsim 2$ GeV becomes 
smaller. Typically, the expectation is that at the SPS (Pb+Pb at 
$\sqrt s=17$ $A$GeV),  the soft component dominates, and at the LHC 
(Pb+Pb at $\sqrt s=5.5$ $A$TeV) the semihard component  is the dominant 
one. At RHIC (Au+Au at $\sqrt s=200$ $A$GeV) one  will be in an 
intermediate  region where both components should be taken into account. 

The semihard processes have also been implemented in several 
event generators for URHIC \cite{MG}. For example, HIJING \cite{HIJING} and 
Parton Cascade Model (and VNI) \cite{GEIGER} rely on the dominance of 
initial semihard particle production at high cms-energies.

In what follows, I am going to focus on how to compute the initial 
minijet and transverse energy production in an $AA$-collisions from 
pQCD, and what are the implications of such a computation for the 
initial conditions of QGP-formation at the LHC and RHIC. I will also 
discuss the uncertainties in the computation.

\section{Production of semihard $g,q,\bar q$ in $AA$}

Hadronic jets originating from high-$p_{\rm T}$ parton-parton scatterings 
have been observed in $p\bar p$ collisions from $p_{\rm T} \gsim 5$ GeV 
 \cite{UA188} up  to $p_{\rm T} \sim 440$ GeV \cite{TEVA}. Below 5 GeV
it becomes very difficult to distinguish the individual transverse energy 
clusters \cite{UA188} from the underlying background, so that minijets 
with $p_{\rm T}\sim 2$ GeV cannot be observed directly \footnote{In this sense 
minijet is not a ``jet'' at all.}. In $AA$ collisions, where hundreds (RHIC) 
or thousands (LHC) of minijets are expected to be produced within the 
central rapidity unit, it becomes certainly impossible to distinguish 
them individually. However, due to their abundance, minijets are expected 
to dramatically contribute to QGP-formation and its further evolution 
at the LHC and RHIC.

Unlike in a jet measurement (or next-to-leading order (NLO) calculation) 
we do not define any jet size for an individual (mini)jet, but rather, 
our goal will now be to compute  minijet and transverse energy production 
in the {\it whole} central rapidity  unit $|y|\le 0.5$. The key quantity 
is the integrated minijet cross section,
\begin{eqnarray}
\nonumber
\sigma_{\rm jet}(\sqrt s, p_0) = 
\frac{1}{2}\int_{p_0^2} dp_{\rm T}^2dy_1dy_2\sum_{{ij<kl>=}\atop{q,\bar q,g}}
        x_1f_{i/A}(x_1,Q) \, x_2f_{j/A}(x_2,Q) \times    \\
     \times  \biggl[ \frac{d\hat\sigma}{d\hat t}^{ij\rightarrow kl}
                 \hspace{-0.7cm}(\hat s, \hat t,\hat u)
     +        \frac{d\hat\sigma}{d\hat t}^{ij\rightarrow kl}
                 \hspace{-0.7cm}(\hat s, \hat u,\hat t)
      \biggr]     \frac{1}{1+\delta_{kl}}.
\,\,\,\,\,\,\,\,\,\,\,\,\,\,\,\,\,\,\,\,\,\,\,\,\,\,\,\,\,\,\,\,\,\,\,\,\,\,\,
\label{sigmajet}
\end{eqnarray}
Collinear factorization is assumed, and $f_{i/A}(x,Q)$ are the 
number densities (per nucleon) of partons $i$ in a nucleus $A$ 
at a fractional momentum 
$x$ and a factorization scale $Q$, chosen here as $Q=p_{\rm T}$. 
From conservation of energy and  momentum in simple 
$2\rightarrow2$ kinematics with negligible initial transverse momenta
it follows that
\begin{equation}
x_1 = \frac{p_{\rm T}}{\sqrt s}({\rm e}^{y_1}+ {\rm e}^{y_2}) 
\,\,\,\,\,\,{\rm and}\,\,\,\,\,\,
x_2 = \frac{p_{\rm T}}{\sqrt s}({\rm e}^{-y_1}+ {\rm e}^{-y_2}),
\end{equation}
where $y_{1,2}$ are the rapidities of the outgoing partons $k,l$ and 
$p_{\rm T}$ is the magnitude of their transverse momentum. Note that in this case
$k$ and $l$ are back-to-back in transverse momentum. 
The kinematic invariants on the parton level become
\begin{eqnarray}
\hat s = x_1x_2s = 2p_{\rm T}^2(1+\cosh Y),\,\,\,
\hat t = -p_{\rm T}^2 (1+{\rm e}^{-Y}),\,\,\,\,\
\hat u = -p_{\rm T}^2 (1+{\rm e}^{Y}),\,\,\,\,\,\,\,\
\end{eqnarray}
where $Y = y_1-y_2$.

In the lowest order pQCD there are eight types of subprocesses 
$\hat \sigma^{ij\rightarrow kl}$:
\begin{eqnarray}
gg \rightarrow gg, q_i\bar q_i\,\,\,\,\,\,\, 
gq \rightarrow gq\,\,\,\,\,\,\, 
q_i \bar q_i \rightarrow q_j \bar q_j, q_i \bar q_i,gg\,\,\,\,\,\,\, 
q_i q_j \rightarrow q_i q_j \,\,\,\,\,\,\,
q_i q_i \rightarrow q_i q_i.\,\,\,\,\,\,\,\,\,\,\,\,\,\,
\label{subxs}
\end{eqnarray}
The cross sections can be found in the literature \cite{XS}.
In $\sigma_{\rm jet}$ the contribution from $gg \rightarrow gg$ 
is clearly dominant. In the computation below, I will use (\ref{subxs}),
but as a little side-remark it is perhaps worth noticing that the 
``single  effective subprocess'' -approximation \cite{SES} works also 
nicely: it is straightforward to show that 
\begin{equation}
\frac{[d\hat\sigma(\hat s,\hat t,\hat u)/d\hat t] ^{gg\rightarrow gg}}
{  [d\hat\sigma(\hat s,\hat t,\hat u)/d\hat t]^{gq\rightarrow gg}
+  [d\hat\sigma(\hat s,\hat u,\hat t)/d\hat t]^{gq\rightarrow gg} }
 = \frac{9}{4} +\left\{ \begin{array}{ll}
                      {\cal O}(\chi), & \chi \ll 1 \\
                      {\cal O}(1/\chi), & \chi \gg 1 
                      \end{array}
              \right.
\end{equation}
where $\chi \equiv \hat u/\hat t$. Considering similarly the process
$qq\rightarrow qq$, one obtains an approximate relation between the 
different  main subprocesses:
\begin{equation}
(gg \rightarrow gg) : (gq \rightarrow gq) : (qq \rightarrow qq)  = 
1 : \frac{9}{4} : (\frac{9}{4})^2,
\end{equation}
so that Eq. (\ref{sigmajet}) becomes simply \cite{EKL}
\begin{equation}
\sigma_{\rm jet}(\sqrt s, p_0) = \frac{1}{2}\int_{p_0^2} dp_{\rm T}^2dy_1dy_2
x_1F(x_1,Q) x_2F(x_2,Q) \frac{d\hat\sigma}{d\hat t}^{gg \rightarrow gg}
\end{equation}
with $F(x,Q)\equiv xg(x,Q) + \frac{4}{9}\sum_q x[q(x,Q) + \bar q(x,Q)]$.

Let us first neglect all nuclear effects in parton densities and take
$f_{i/A}=f_{i/p}\equiv f_i$, which is a reasonable first approximation to start with.
Then, from the $pp$-level quantity $\sigma_{\rm jet}$, we obtain  the 
{\it average} number of produced minijets with $p_{\rm T}\ge p_0$ in an $AA$ 
collision with an impact parameter ${\bf b}$ simply by multiplying
$\sigma_{\rm jet}$ by the nuclear overlap function $T_{AA}$ \cite{KLL,EKL}: 
$\bar N_{AA}({\bf b}) = 2 T_{AA}({\bf b}) \sigma_{\rm jet}$, where
\begin{equation}
T_{AA}({\bf b}) = \int d^2 {\bf s} \,T_A({\bf s}) T_A({\bf b-s}),  
\end{equation}
where the thickness function is
\begin{equation}
T_A({\bf s}) = \int dz \,n_A(\sqrt{s^2+z^2}) 
\end{equation}
with normalizations 
$\int d^2{\bf b} T_{AA}({\bf b}) = A^2$  and 
$\int d^2{\bf s} T_A({\bf s}) = A,$ {\it i.e.} the hard cross sections 
in $AA$ scale as $A^2$, as expected in absence of any nuclear effects 
in $f_{i/A}$. For central collisions with Woods-Saxon nuclear densities 
$n_A$ we get $T_{AA}(b=0) \approx A^2/(\pi R_A^2)$, so clearly 
$N_{AA} \sim A^{4/3}$. Here one should notice that this approach is based 
on {\it independent} binary parton collisions, and is valid when each 
semihard sub-collision consumes only a  negligible fraction of the 
beam-energy,  and when there are  sufficiently many partons available 
for scattering. These conditions are  fulfilled for minijet production 
at central  rapidities even  up to the LHC-energies and $A\sim 200$. 
However, at the LHC for large $y$ together with a very large $A$, 
one eventually  runs into trouble with energy and baryon number 
conservation \cite{EK96}, which is an indication that  in this 
regime one should consider coherence effects in parton production. 
This is easy to understand simply by  noticing that the available 
energy  and valence quark number scale as $\sim A$, while 
$N_{AA}\sim A^{4/3}$. For RHIC energies, however, 
there does not seem to be a problem in the above approach for any $y$.
 \cite{EK96}
 
We want to focus on the central rapidity unit, so we will have to do 
some book-keeping of where the partons are produced in rapidity, and define 
our acceptance cuts. Especially, we want to compute the average 
transverse energy carried by the partons in the central rapidity unit. 
Furthermore, we would like to keep track of contributions from gluons, 
quarks and antiquarks separately. Let us therefore define the 
$E_{\rm T}$-distribution for each flavour $f$ in our acceptance window 
$\Delta y$ as \cite{EKL,EK96}:
\begin{eqnarray}
\frac{d\sigma^f}{dE_{\rm T}} = 
\frac{1}{2}\int dp_{\rm T}^2dy_1dy_2\, \sum_{ij\atop {\langle kl\rangle}}  
x_1f_i(x_1,Q)\, x_2 f_j(x_2,Q)\frac{1}{1+\delta_{kl}}
\times\hspace{3cm}\nonumber \\
\times\biggl\{
\frac{d\hat\sigma}{d\hat t}^{ij\rightarrow kl} \hspace{-0.7cm}(\hat s, \hat t,\hat u)
\,\delta(E_{\rm T}-
[\delta_{fk}\epsilon(y_1) + \delta_{fl}\epsilon(y_2)]p_{\rm T})
+\nonumber
\,\,\,\,\,\,\,\,\,\,\,\,\,\,\,\,\,\,\,\,\,\,\,\,\,\,\,\,\,\,\,\,\,\,\,\,\,\,\,
\\
+ \frac{d\hat\sigma}{d\hat t}^{ij\rightarrow kl}\hspace{-0.7cm}(\hat s, \hat u,\hat t)
\,\delta(E_{\rm T}-
[\delta_{fl}\epsilon(y_1) + \delta_{fk}\epsilon(y_2)]p_{\rm T})
\biggr\},
\,\,\,\,\,\,\,\,\,\,\,\,\,\,\,\,\,\,\,\,\,\,\,\,\,\,\,\,\,\,\,\,\,\,\,\,\,\,\,
\label{dsdet}
\end{eqnarray}
with a normalization
$
\sum_f \int dE_{\rm T}\frac{d\sigma^f}{dE_{\rm T}} = 
\sigma_{\rm jet}(\sqrt s,p_0). 
$
Above, our acceptance is defined through a ``measurement function'' 
$\epsilon(y)$, which will in the following be chosen as a step function
\begin{equation}
\epsilon(y) = \left\{ \begin{array}{ll}
                      1, & \mbox{if $|y|\le0.5$} \\
                      0, & \mbox{otherwise.}
                      \end{array}
              \right.
\end{equation}
\begin{figure}[tb]
\vspace*{6.0cm}
\centerline{\epsfxsize=10.0cm\epsfbox{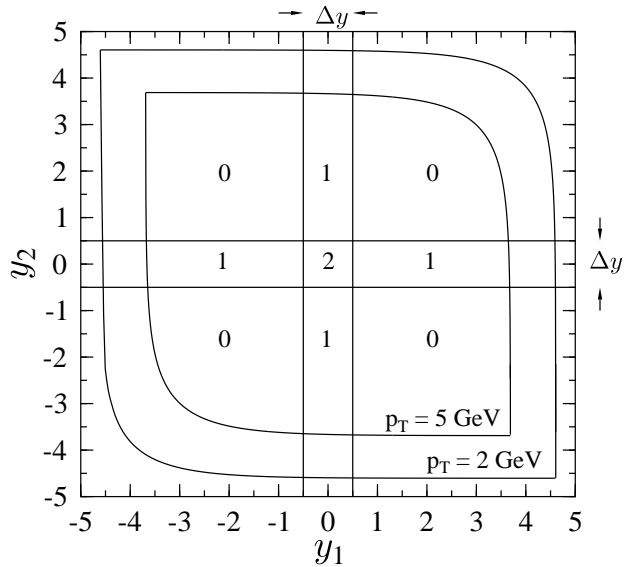}}
\vspace*{-7.5cm}
\caption{{\footnotesize  
Kinematic cuts in rapidity.  The oval regions  show the rapidity  
phase space (\ref{yPS}) for fixed $p_{\rm T} = 2$ GeV and 
5 GeV, and  $\sqrt s=200$ GeV.  The values of $\epsilon(y_1)+\epsilon(y_2)$ 
corresponding to our acceptance region, $\Delta y:|y|\le 0.5$, are indicated.
}}
\label{ycuts}
\end{figure}

In Fig. 1, the integration region in the $(y_1,y_2)$-plane 
(with a fixed $p_{\rm T}$),
\begin{eqnarray}
|y_1|\le {\rm log} \biggl[\frac{\sqrt s}{2p_{\rm T}} + 
\sqrt{(\frac{\sqrt s}{2p_{\rm T}})^2 -1}\,\,\biggr],
\,
-{\rm log}\biggl[\frac{\sqrt s}{p_{\rm T}}-{\rm e}^{-y_1} \biggr]
\le y_2 \le \log\biggl[\frac{\sqrt s}{p_{\rm T}}-{\rm e}^{y_1}\biggr],
\hspace{1.3cm}
\label{yPS}
\end{eqnarray}
is illustrated together with the cuts corresponding to our 
acceptance $\Delta y:\,|y|\le 0.5$. After this, it is straightforward 
to compute the first $E_{\rm T}$-moment for 
each flavour $f$  in the semihard collision, 
\begin{eqnarray}
\nonumber
\sigma_{\rm jet}(\sqrt s,p_0)\langle E_{\rm T}^f\rangle_{\Delta y} = 
\int dE_{\rm T}E_{\rm T} \frac{d\sigma^f}{dE_{\rm T}} \hspace{6.3cm}\\
\nonumber
= \int dp_{\rm T}^2dy_1dy_2\, \sum_{ij\atop {\langle kl\rangle}}  
x_1f_i(x_1,Q)\, x_2 f_j(x_2,Q)\times\hspace{1.6cm}
\\
\times \frac{1}{1+\delta_{kl}}\biggl[
\delta_{fk}\frac{d\hat\sigma}{d\hat t}^{ij\rightarrow kl}
\hspace{-0.7cm}(\hat s,\hat t,\hat u)
+
\delta_{fl}\frac{d\hat\sigma}{d\hat t}^{ij\rightarrow kl}
\hspace{-0.7cm}(\hat s, \hat u,\hat t)
\biggr] p_{\rm T}\epsilon(y_1)
\hspace{0.8cm}
\label{sET}
\end{eqnarray}

In an $AA$ collision with an impact parameter ${\bf b}$,
the average transverse energy carried by semihard partons of flavour $f$ 
to the acceptance region becomes then \cite{EKL,EK96}
\begin{equation}
\bar E_{\rm T}^f({\bf b},\sqrt s,p_0,\Delta y) = 
T_{AA}({\bf b}) \sigma_{\rm jet}(\sqrt s,p_0)
\langle E_{\rm T}^f\rangle_{\Delta y},
\label{ETAA}
\end{equation}
where $T_{AA}({\bf b})\sigma_{\rm jet}(\sqrt s,p_0)$ is the 
average number of semihard collisions (all $y$) and 
$\langle E_{\rm T}^f\rangle_{\Delta y}$ is the average transverse energy 
carried by the flavour $f$ at $|y|\le 0.5$ in each semihard collision.
 
We can also compute the number distribution $d\sigma^f/dn$ for each flavour
in  a similar way  (in Eq.(\ref{dsdet}) replace  $E_{\rm T}$ by $n$, 
and in the delta-functions $p_{\rm T}$ by $1$) to obtain the average 
number of  flavour $f$ produced within our acceptance window:
\begin{equation}
\bar N_{AA}^f({\bf b},\sqrt s,p_0,\Delta y) = 
T_{AA}({\bf b}) \sigma_{\rm jet}(\sqrt s,p_0)\langle n^f\rangle_{\Delta y},
\label{NAA}
\end{equation}
where $\sigma_{\rm jet}(\sqrt s,p_0)\langle n^f\rangle_{\Delta y}$ 
can be obtained directly from  Eq. (\ref{sET}) simply by removing 
the weight $p_{\rm T}$. The normalization 
$\sum_f\sigma_{\rm jet}(\sqrt s,p_0)
\langle n^f\rangle_{\Delta y\rightarrow\infty} 
= 2\sigma_{\rm jet}(\sqrt s,p_0)$ 
can be verified from Eqs. (\ref{sigmajet}) and (\ref{sET}).

\section{Initial conditions at $\tau = 0.1$ fm/$c$}

The subprocess cross sections $d\hat \sigma^{ij}/d\hat t$ diverge as 
$\sim p_{\rm T}^{-4}$ when $p_{\rm T}\rightarrow 0$, which is why we cannot 
reliably extend our computation below $p_0 = 1...2$ GeV. The 
QCD-evolution of the parton densities does not improve the situation enough 
because of two competing effects:  towards smaller $p_T$ both $x$ and $Q$ 
decrease; the downwards QCD-evolution  reduces the densities (for fixed, 
small $x$) but the simultaneous small-$x$ increase of the densities has an 
effect in the other direction. We would like to 
extend our computation down to as low values of $p_{\rm T}$ as possible 
but making sure we can still believe that we are dealing with 
perturbative quantities. Of course, only after computing the NLO minijet 
cross sections, the question of convergence of the perturbation series 
can be better answered. For large-$p_{\rm T}$ jet measurements the 
NLO-calculation  has been done \cite{EKS} but for the few GeV 
$p_{\rm T}$-range and for the rapidity acceptance we have in mind this 
has not been completed yet.

So, how to fix the parameter $p_0$? 
In Fig. 2, I have plotted the behaviour of 
$\sigma_{\rm jet}(\sqrt s,p_0)\langle E_{\rm T}^f\rangle_{\Delta y}$ and 
$\sigma_{\rm jet}(\sqrt s,p_0)\langle n^f\rangle_{\Delta y}$ for the LHC 
and RHIC energies as functions of $p_0$. In the computation, I have assumed 
that $f_{i/A} \approx f_i$.  I have used the GRV94 LO-distributions  \cite{GRV94}, which 
reproduce the observed small-$x$ rise of the partons densities reasonably 
well. Note that there is no {\it ad hoc} ``$K$-factor'' included in the 
computation.

One way to fix $p_0$ is to formulate a saturation criterion
 \cite{BM87,GLRpr,EK96} for the produced parton system in URHIC: 
let us assign a transverse area $\pi/p_{\rm T}^2$ for each parton within
 our rapidity acceptance. Most of the partons have $p_{\rm T}\sim p_0$, 
so that the effective total transverse area occupied by the partons is 
$\bar N_{AA}\times\pi/p_0^2$. The available nuclear transverse area can 
be estimated as $\pi R_A^2$, so that the saturation of parton production 
in $|y|\le 0.5$ should occur when 
$\bar N_{AA}\times\pi/p_0^2 \gsim \pi R_A^2$, {\it i.e.} when
\begin{equation}
\sigma_{\rm jet}(\sqrt s,p_0)\langle n^f\rangle_{|y|\le0.5} \gsim
134 \biggl({p_0\over 2\,{\rm GeV}}
\biggr)^2 \,{\rm mb},
\label{saturation}
\end{equation}
where $R_{\rm Pb} = 6.54$ fm and $T_{\rm Pb Pb}(0) = 32/{\rm mb}$ has been 
used. This procedure results in $p_0\approx 2$ GeV for the LHC, and 
$p_0\approx 1$ GeV for RHIC, as illustrated in Fig. 2. 

Saturation of parton production can also be studied in a more 
self-consistent - but still phenomenological - manner by computing a 
screening mass (electric, static) generated  by the partons produced 
by a fixed time $\tau<1/p_{\rm T}$ \cite{EMW}. This mass then screens further 
parton production at smaller $p_{\rm T}$ and later in time.
The more partons are produced the stronger the screening becomes, and, the 
more damped further parton production is. In this way, saturation 
of transverse energy production seems again to take place at 
$p_{\rm T}\sim 2$ GeV for LHC and  $p_{\rm T}\sim 1$ GeV for RHIC \cite{EMW}.

Lower limits for $p_0$ may also be studied in an eikonal 
approach \cite{EIKONAL}, or,
by convoluting the parton cross sections with the fragmentation functions 
of each parton flavour into pions and kaons \cite{fragm}. Comparison of the 
fragmentation function approach with the measured $p_{\rm T}$-distributions 
of charged particles in $p\bar p$  collisions is successful 
at $\sqrt s\ge 200$ 
GeV at large $p_{\rm T}$ \cite{borzumati}, whereas it seems that allowing 
parton scatterings with $p_{\rm T}<1.5$ GeV would result in an overestimate of 
the charged particle  $p_{\rm T}$-distributions.  Therefore, let me fix 
$p_0 = 2$ GeV both for the LHC and for RHIC. Estimates for smaller $p_0$
can be easily obtained from Fig. 2.

\begin{figure}[hb]
\vspace*{4cm}
\centerline{\epsfxsize=10.0cm\epsfbox{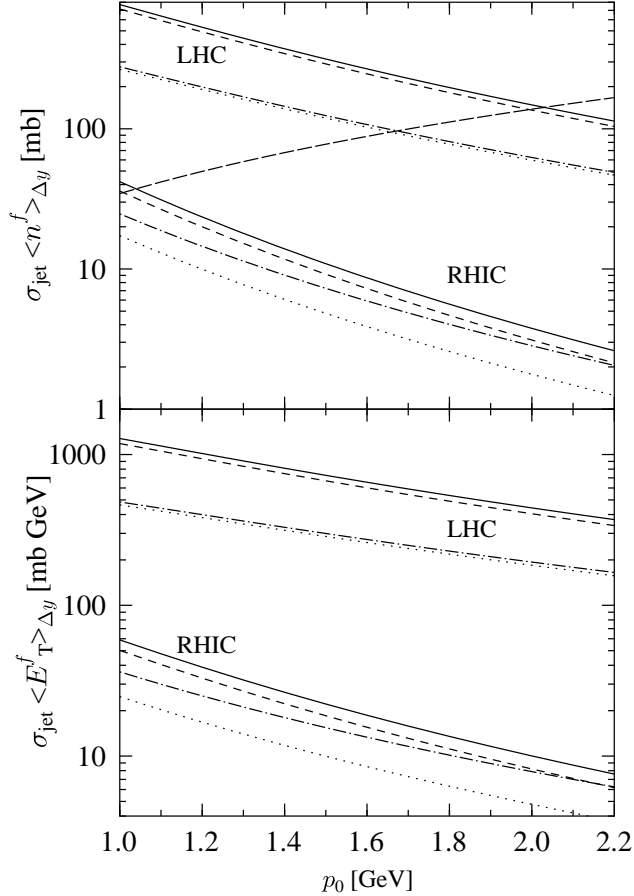}}
\vspace*{-2.3cm}
\caption{{\footnotesize quantities 
$\sigma_{\rm jet}(\sqrt s,p_0)\langle n^f\rangle_{\Delta y}$ and 
$\sigma_{\rm jet}(\sqrt s,p_0)\langle E_{\rm T}^f\rangle_{\Delta y}$ from 
Eqs. (\ref{sET}-\ref{NAA}) for $\sqrt s=5500$ GeV (LHC) and   
$\sqrt s=200$ GeV (RHIC) as functions of $p_0$ for $\Delta y: |y|\le 0.5$. 
The solid curves show the total results, the dashed ones the gluon 
contributions. For the figure, quark and antiquark contributions have been
multiplied by a factor 10 (7) for the LHC (RHIC) and they are shown with 
dot-dashed and dotted curves, respectively. The saturation limit 
from Eq. (\ref{saturation}) is the increasing dashed line in the upper panel.  
}}
\label{cuts}
\end{figure}

Table \ref{table} shows the average numbers and transverse energies of 
semihard partons at $\tau=0.1$ fm$/c$ with $|y|\le 0.5$ and 
$p_{\rm T}\ge 2$ GeV in  central Pb--Pb collisions at the LHC and RHIC, 
as given by  Eqs. (\ref{NAA}) and (\ref{ETAA}).

\begin{table}
\center
\begin{tabular}{|c|c|c|c|c|c|}
\hline
      & $|y|<0.5$                    & total & $g$   & $q$    & $\bar q$   \\
\hline
LHC:  & $\bar N_{\rm PbPb}^f$          &4741   & 4350  & 200  & 191      \\
      & $\bar E_{\rm T}^f$  &14160  & 12950 & 619  & 590       \\
 
\hline 
RHIC: & $\bar N_{\rm PbPb}^f$          & 121  & 99.6 &13.0  & 8.10     \\
       & $\bar E_{\rm T}^f$ & 321  & 263 &36.1  & 21.9    \\
 
\hline
\end{tabular}
\vspace{0.5cm}
\caption[1]{ {\footnotesize The initial conditions at $\tau = 0.1$ fm:  
$N_{\rm PbPb}^f$ and $\bar E_{\rm T}^f$ for $b=0$, $\sqrt s/A = 5500$ 
GeV (LHC) and  200  GeV (RHIC), $p_0 = 2$ GeV and $|y|\le0.5$. }}
\label{table}
\end{table}

We make the following five interesting observations:

1. Gluons clearly outnumber quarks and antiquarks and dominate  
$E_{\rm T}$-production at $\tau \sim 0.1$ fm/$c$ both at the LHC and RHIC. 
The QGP is actually gluon plasma  to a first approximation.

2. With $p_0 = 2$ GeV $\gg \Lambda_{\rm QCD}$  the glue is saturated 
at the LHC. This indicates that the pQCD-domain is dominant for 
$E_{\rm T}$-production.  In other words, more gluons are produced 
into the system at $p<p_0$ but  their transverse energy is relatively 
small. For RHIC I expect the soft component to be still important.

3. Estimating the volume of the system at $\tau_{\rm h}=0.1$ fm/$c$ and with 
$\Delta y=1$ by  $ V_{\rm i}=\pi R_{\rm Pb}^2\Delta y \tau_{\rm h}$, 
we can convert the results into local energy and number densities:
$\epsilon_f^{\rm pQCD} = \bar E_{\rm T}^f/V_{\rm i}$ and 
$n_f^{\rm pQCD} = \bar N_{\rm PbPb}^f/V_{\rm i}$. 
From the numbers in Table \ref{table} for the LHC, we see that 
$E_{\rm T}/$gluon $\approx 3$ GeV (naturally, since $p_0=2$ GeV). 
On the other hand, for an ideal gas of massless bosons
 $\epsilon_g^{\rm ideal}/n_g^{\rm ideal}\approx 2.7 T_{\rm eq}$. 
For an ideal gas with $\epsilon_g^{\rm ideal} = \epsilon_g^{\rm pQCD}$ 
we get $T_{\rm eq}\approx 1.1$ GeV, which means that
$(\epsilon_g/ n_g)^{\rm pQCD}\approx(\epsilon_g/n_g)^{\rm ideal}$,
and the glue seems to be initially thermalized at the LHC. \cite{EK96} 
This result is essentially due to the small-$x$ increase of the 
gluon distributions \cite{EKR}. At RHIC the gluon thermalization will 
take place a little later. One should also keep in mind that a 
complete thermalization requires  locally isotropic momentum 
distributions, and to study this more detailed modelling is needed.

4. Since $n_q, n_{\bar q} \ll n_g$, quarks and antiquarks are initially 
obviously far away from chemical equilibrium both at RHIC and the LHC.

5. By assigning a baryon number $1/3(-1/3)$ for each quark(antiquark) produced
at $\tau=0.1 $fm/$c$, we observe that the initial net baryon number density 
at $|y|\le0.5$ becomes $n_{B-\bar B} = 0.21\,{\rm fm}^{-3}$ for the LHC 
and 0.12  fm$^{-3}$ for RHIC. This density will naturally dilute quite fast 
($\sim 1/\tau$) but it is interesting that for the LHC the initial number is 
actually larger than for RHIC, and even larger than the nuclear matter 
density 0.17 fm$^{-3}$. This behaviour is again due to the small-$x$ 
increase of the gluon densities. Also, for the LHC, it is worth noticing that
the transit time of the colliding nuclei is 
$2R_A/\gamma\sim 0.005 $ fm/$c \ll 0.1$ fm$/c$. 
Finally, the initial net baryon number-to-entropy ratio can also be estimated 
 \cite{EK96}, and we find that $(B-\bar B)/S\sim 1/5000\, (1/2000)$ for the LHC 
(RHIC), so that we are far away from the conditions of the early 
Universe, where the inverse of the specific entropy is $\sim 10^{-9}$.

\section{Discussion and outlook}
I have shown how to compute initial conditions for QGP-production
in URHIC at the LHC and RHIC from the lowest order pQCD. The analysis 
can obviously be sharpened  in various ways: 
by computing the transverse energy production in NLO in $\alpha_s$ 
and by including nuclear effects in the parton distributions in 
a consistent, scale-dependent manner \cite{KJE}. Resummation similar 
to the Drell-Yan case \cite{AEGM}, if applicable, would hopefully 
help to reduce the uncertainty due to the parameter $p_0$. Also 
higher twist effects and more coherent scattering should be 
studied and formulated in more detail at large $y$. Connection 
to the BFKL-physics potentially relevant to the minijets involving 
small-$x$ should be understood better  \cite{ELR} as well. Recently, 
a novel approach based on classical gluon fields has been developed 
 \cite{McL}  for semihard parton production. Applicability of this model 
at the energies of RHIC and the LHC has been studied lately \cite{GMcL}. 
Also, understanding the pre-thermal evolution of the 
newly produced QGP is a challenging task, involving space-time 
dependent phenomena in gauge theories.

To conclude, it is evident that there are several interesting open questions
in pre-thermalization physics of URHIC, and much more precision-work is needed.
Understanding the primary parton production mechanisms will be essential 
for understanding the further evolution and signals of the QGP, 
and more global variables measured at nuclear collisions at the LHC 
and RHIC.

\bigskip
{\bf Acknowledgements.} I would like to thank K. Kajantie for collaboration 
and helpful discussions.

\end{document}